\documentstyle[12pt]{article}
\begin{document}
\newcommand{\be}{\begin{equation}}
\newcommand{\ee}{\end{equation}}
\newcommand{\bea}{\begin{eqnarray}}
\newcommand{\eea}{\end{eqnarray}}
\newcommand{\beas}{\begin{eqnarray*}}
\newcommand{\eeas}{\end{eqnarray*}}

\baselineskip 14 pt
\parskip 12 pt

\begin{titlepage}
\begin{flushright}
{\small hep-th/0405042}\\
\end{flushright}

\begin{center}

\vspace{2mm}

{\Large \bf Charged black holes from near extremal black holes}

\vspace{3mm}

 Gilad Lifschytz

\vspace{1mm}

{\small \sl Department of Mathematics and Physics and CCMSC} \\
{\small \sl University of Haifa at Oranim, Tivon 36006, Israel} \\
{\small \tt giladl@research.haifa.ac.il}

\end{center}

\vskip 0.3 cm

\noindent
We recover the properties of a wide class of far from extremal
 charged black branes from
properties of near extremal black branes,
 generalizing the results of Danielsson, Guijosa and Kruczenski.

\end{titlepage}

\section{Introduction}
The underlying physics of black holes have been the target of many
investigations.  It has been realized that certain near extremal black
holes can be modeled \cite{peet} by a large number of D-branes with
excitations on them. This spurred  the description of string theory in
certain backgrounds  by a large $N$ gauge theory in the strong 't Hooft coupling regime \cite{malda,imsy}. It seemed that some new physics has to come in, in order to describe the far from extremal and Schwarzschild black holes.

Recently generalizing the construction in \cite{dgk} a possibility of 
understanding Schwarzschild black holes in various dimensions \cite{sp,bl,ka}and
rotating neutral black holes \cite{ghm,sp} in various dimensions from the physics
of near extremal black hole have emerged. Since the physics of near
extremal black holes is governed by large N gauge theories, this means
the the physics of far from extremal black holes can be similarly
understood. In particular the understanding of the relationship between area and entropy \cite{ikll1,ikll2}, and
the origin of the horizon size \cite{ikll}
for the near extremal case , goes over to the Schwarzschild black brane case as well.
In this note we generalize this description to the far from extremal
charged black holes case (or black branes). The example of the single
charged black brane in seven dimensions was constructed in
\cite{dgk}\footnote{See also \cite{hms}},
and we generalize it here to any dimension and to the multi charged case.

\section{Single charged black brane}
We start with the simplest case where the black branes carry only one
kind of charge from the ten dimensional perspective. i.e  they are made
from configurations with different numbers of D-branes and anti
D-branes, of the same dimensionality.

The mass, entropy and charge of the non extremal charged p-brane in $D$-dimensions we
wish to understand are given by a two parameter family \cite{dlp} ($D=p+d+3$)
\begin{eqnarray}
M_{bh}=\frac{\omega_{d+1}}{2\kappa^2}L^{p}\mu^{d}(d+1+d\sinh
^{2}\gamma)\nonumber \\
S_{bh}=\frac{2\pi}{\kappa^2}\omega_{d+1}L^{p}\mu^{d+1}\cosh \gamma \nonumber \\
Q_{bh}=\frac{1}{2\sqrt{2}\kappa}\omega_{d+1}L^{p}\mu^{d}\sinh 2\gamma
\label{farext}
\end{eqnarray}

The near extremal entropy of these black holes is given by \cite{kt}
\begin{eqnarray}
S=a(\frac{E}{2})^{\lambda}\sqrt{M_{p,0}}\nonumber \\
\lambda=\frac{d+1}{d}-\frac{1}{2}\ \ ,\  D=p+d+3\nonumber \\
a=2^{\frac{1}{d}+2}\pi \omega_{d+1}^{-\frac{1}{d}}d^{-\frac{d+1}{d}}
\lambda^{-\lambda}L^{-\frac{p}{d}}\kappa^{\frac{2}{d}}\nonumber \\
M_{p,0}=\frac{q_{p}}{\sqrt{2}\kappa}L^{p}
\label{nearext}
\end{eqnarray}
where $q_{p}$ is the charge per unit volume and $E$ is the energy above extremality.

We will assume that the thermodynamics of the brane and anti brane
system is unaffected by the presence of the other, and is controlled by
the thermodynamics of the near extremal branes, even when $E$ is
large. 
One can then try and
maximise the entropy of the brane anti-brane system to read off the parameters of the system. If we do this, one condition is 
that the temperature of the excitations on the brane is equal to the temperature of the excitations on the anti brane leading to an incorrect result. Now usually this is a natural condition, when two gases of excitation are in thermodaynamical contact. This however is not the case here. 
 The stability of the system means that the
strings connecting the brane and anti-brane are actually massive and
since the temperature on the branes and anti branes will later seen to
be small this means that the two gases of excitations on the brane and
anti brane are effectively (after stabilization) decoupled. What seems
to work is requiring that the energy of the gas of excitation on the
brane and anti-brane be equal
(in the Schwarzschild case this is the same as equall temperatures). 
We believe this condition arises from the 
condition of stability of the system, but can not currently derive
this. 
Note however that this indicates that the energy needed for
stabilization is a function of the number of tachyonic modes, which is
equall on both configurations.

We will thus just proceed with this assumption.
Let us label the mass of the branes (anti-branes) by $M_{p,0}(M_{\bar{p},0})$, and the energy of excitation on both branes by $E$.
Then we write
\begin{eqnarray}
M_{f}=M_{p,0}+M_{\bar{p},0}+E\nonumber \\
S_{f}=a(\frac{E}{2})^{\lambda}(\sqrt{M_{p,0}}+\sqrt{M_{\bar{p},0}})\nonumber \\
Q_{f}=\frac{\sqrt{2}\kappa}{L^{p}}(M_{p,0}-M_{\bar{p},0})
\label{field}
\end{eqnarray}
We now maximise the entropy with respect to $M_{p,0}$ (basically the
number of branes) while keeping the mass and charge fixed. This gives
the relationship
\begin{equation}
E=4\lambda\sqrt{M_{p,0}M_{\bar{p},0}}
\label{em}
\end{equation}
We can plug this back to the equation for the mass and charge 
and then solve for
$M_{p,0}$.
If the mass and charge are given by equation (\ref{farext}) then the solution of
maximising the entropy gives
\begin{eqnarray}
M_{p,0}=\frac{\omega_{d+1}}{2\kappa^2}L^{p}\mu^{d}\frac{d}{4}e^{2\gamma}\nonumber \\
M_{\bar{p},0}=\frac{\omega_{d+1}}{2\kappa^2}L^{p}\mu^{d}\frac{d}{4}
e^{-2\gamma}\nonumber \\
E=\lambda\frac{\omega_{d+1}}{2\kappa^2}L^{p}\mu^{d}d
\label{sol1}
\end{eqnarray}
We can now plug these expression into the expressions of the entropy
in equation (\ref{field}) and we find
\begin{equation}
S_{f}=2^{-\lambda}S_{bh}
\end{equation}
So up to a proportionality constant for the entropy we have reproduced
the dependence of the entropy on the charge and mass  of the far from
extremal charged black brane.

\section{Multi brane black branes}

In this section we will generalize the above construction to the
situation where there are more than one type of branes involved, but
in some lower D-dimension it can be represented as a single charged
black brane.
We want to reproduce the properties of the black branes discussed in
\cite{dlp} (for a general $N$),
\begin{eqnarray}
M_{bh}=\frac{\omega_{d+1}}{2\kappa^2}L^{p}\mu^{d}(d+1+dN\sinh ^{2}\gamma)\nonumber \\
S_{bh}=\frac{2\pi}{\kappa^2}\omega_{d+1}L^{p}\mu^{d+1}\cosh^{N} \gamma \nonumber \\
Q_{bh}=\frac{\sqrt{N}}{2\sqrt{2}\kappa}\omega_{d+1}L^{p}\mu^{d}\sinh 2\gamma
\label{farext1}
\end{eqnarray}
The near extremal entropy is given by \cite{kt}
\begin{eqnarray}
S(E) &=& b' q_{p}^{\frac{N}{2}}E^{\lambda}\nonumber \\
b' &=& 2^{\frac{1}{d}-\frac{N}{4}+2}\pi
\omega_{d+1}^{-\frac{1}{d}}d^{-\frac{d+1}{d}}\lambda^{-\lambda}N^{-\frac{N}{4}}
L^{p(1-\lambda)}\kappa^{\frac{2}{d}-\frac{N}{2}}\nonumber \\
M_{p,0}& = &\frac{q_{p}\sqrt{N}}{\sqrt{2}\kappa}L^{p}\ \ , \ \lambda=\frac{d+1}{d}-\frac{N}{2}
\end{eqnarray}
Where $N$ represents the number of diffrent D-branes. 
Now there is a question what is the near extremal entropy of an object
made out of the black brane and anti black brane. When a single brane
was involved the brane anti brane is given by the equation in
(\ref{field}), but now things are different. To see this let us
imagine we are considering the configuration of the
$D1-D5$-branes. The low energy states are given by the $D1-D5$ strings
while the $D1-D1$ and $D5-D5$ are presumed massive. If we now take a
system with a $D1-D5$ and an $\bar{D}1-\bar{D}5$, there are four
sectors of low energy excitations, which include also the
$\bar{D}1-D5$ and the $D1-\bar{D}5$ strings. If we consider the
situation in which in six dimensions the charge of the $D1$ is equal
the charge of the $D5$ then the near extremal mass and entropy of the
system can be written as
\begin{eqnarray}
M_{f}=M_{p,0}+M_{\bar{p},0}+E\nonumber \\
S_{f}=a(\frac{E}{2^{N}})^{\lambda}(\sqrt{M_{p,0}}+\sqrt{M_{\bar{p},0}})^{N}\nonumber \\
a=b'(\frac{\sqrt{2}\kappa}{L^{p}})^{N/2}\nonumber \\
Q_{f}=\frac{\sqrt{2}\kappa}{\sqrt{N}L^{p}}(M_{p,0}-M_{\bar{p},0}).
\label{field1}
\end{eqnarray}
Where the energy was equally distributed among all $2^{N}$ kinds of low
energy excitations.
We now can maximise the entropy with respect to $M_{p,0}$ keeping the
mass and charge fixed, and we find the condition,
\begin{equation}
E=\frac{4\lambda}{N}\sqrt{M_{p,0}M_{\bar{p},0}}
\end{equation}
We can now again plug into the mass and charge to find
\begin{eqnarray}
M_{p,0}=\frac{\omega_{d+1}}{2\kappa^2}L^{p}\mu^{d}\frac{Nd}{4}e^{2\gamma}\nonumber \\
M_{\bar{p},0}=\frac{\omega_{d+1}}{2\kappa^2}L^{p}\mu^{d}\frac{Nd}{4}
e^{-2\gamma}\nonumber \\
E=\lambda\frac{\omega_{d+1}}{2\kappa^2}L^{p}\mu^{d}d
\label{sol2}
\end{eqnarray}
This can now be plugged into the expressions for entropy 
equation (\ref{field1}) and we find
\begin{equation}
S_{f}=2^{-N\lambda}S_{bh}.
\end{equation}
Again we reproduce the entropy mass (and charge) relationship up to a numerical constant.

\section{Multi charged black holes}
We now come to a more general situation where the different D-branes
carry diffrent charges. The black holes are described in \cite{ct}
and their mass and entropy  (upon some re labeling) are given by
\begin{eqnarray}
M_{bh}=\frac{b}{2}\mu^{D-3}(\sum_{i=1}^{N}\cosh \gamma_{i} +2\lambda)\nonumber \\
b=\frac{\omega_{D-2}}{2\kappa^2}(D-3)V_{p}\ \ \ ,\ 
\lambda=\frac{D-2}{D-3}-\frac{N}{2}\nonumber \\
S_{bh}=c\mu^{D-2}\Pi_{i=1}^{N}\cosh \gamma_{i}\nonumber \\
c=\frac{4\pi b}{D-3}.
\label{farext2}
\end{eqnarray}

The near extremal entropy is given by \cite{kt1}
\begin{eqnarray}
S=\tilde{a}E^{\lambda}\Pi_{i=1}^{N}\sqrt{M_{p_{i},0}}.\nonumber \\
\tilde{a}=c(b\lambda)^{-\lambda}b^{-N/2}\nonumber \\
M_{p,0}=bq_{p}
\end{eqnarray}
Following the same logic as before we can try and model the far from
extremal black hole using the near extremal relationship. After taking
into account all configurations of brane that contribute to the
entropy, we can write,
\begin{eqnarray}
M_{f}=\sum_{i}^{N}(M_{p_{i},0}+M_{\bar{p}_{i},0})+E \nonumber \\
S_{f}=\tilde{a}(\frac{E}{2^{N}})^{\lambda}\Pi_{i=1}^{N}(\sqrt{M_{p_{i},0}}+\sqrt{M_{\bar{p}_{i},0})}.
\label{field2}
\end{eqnarray}
Maximizing the entropy for fixed charge and mass gives the
relationship
\begin{equation}
E=4\lambda \sqrt{M_{p_{i},0}M_{\bar{p}_{i},0}},
\end{equation}
for all $i$.

Solving the equations gives
\begin{eqnarray}
M_{p_{i},0}=\frac{b}{4}\mu^{D-3} e^{2\gamma_{i}}\nonumber \\
M_{\bar{p},0}=\frac{b}{4}\mu^{D-3} e^{-2\gamma_{i}}\nonumber \\
E= b\lambda \mu^{D-3} 
\label{sol3}
\end{eqnarray}
Plugging it back to the expression for the entropy equation
(\ref{field2})
gives the entropy of the far from extremal charged black hole,
equation 
(\ref{farext2}), up to a
numerical factor of $2^{-\lambda N}$. Note that in all cases, 
as noticed in \cite{dgk}, the
entropy would be exactly the correct one if the energy on each brane
configuration would have been the total energy.

\section{Horizon and temperature}

In this section we would like to briefly discuss some physical
properties of the brane system. If we look at equations
(\ref{sol1}),(\ref{sol2}) and (\ref{sol3}), we see
that the dependence of $E$  on the parameter of the solutions (i.e $\mu$ and
$\gamma$) $ E \sim \mu^{d}$ (or $\mu^{D-3}$) are exactly the same as the dependence of the energy of the
near extremal black hole on these parameters \cite{kt,kt1}. Now since in the
coordinate system used in \cite{kt,kt1} $\mu$ is also the radius of the
horizon, this means that the emergence of a large horizon size has its
origin in the same physical phenomena as the emergence of the near
extremal horizon . Thus again as in the Schwarzschild case the charged black brane horizon can be understood as the regulated transverse size of the gauge theory state\cite{susskind,ikll}, and can be 
seen by a brane probe as the locus of the tachyonic instability region \cite{kl,kl1}.

While the size of the state given by the branes or anti branes is the same
(it is only a function of the energy of the excitations on them) the temperature is different. The relationship between the temperature of the black brane and the field theory temperature on each brane configuration just follows from the regular thermodynamical relationship
\begin{equation}
\frac{1}{T_{bh}}=\frac{\partial S}{\partial M}=\frac{\partial S}
{\partial M_{p_{i},0}}\frac{\partial M_{p_{i},0}}{\partial M}+
\frac{\partial S}{\partial E}\frac{\partial E}{\partial M}=
2^{-N}\sum_{i=1}^{N}\frac{1}{T_{i}}.
\end{equation}
Where in the last equality we have used that since the entropy was
maximal,
\begin{equation}
2\frac{\partial S}{\partial E}=\frac{\partial S}{\partial
  M_{p_{i},0}}.
\end{equation}
The appearance of more that one temperature should have some signal
in the supergravity properties of these black holes, at least through the 
properties of the quasinormal mode.
The temperatures on each brane configuration should be low for the
approximation we have been using to be justified. This means that we
should be quite far from extremality ($\gamma_{i}$'s should not be too
big).
As one approaches extremality at least one of the temperatures becomes
high, and in particular can become high enough to excite the massive
modes between brane and anti-brane, this can lead to the
disappearance of the branes or anti-branes through annihilation.

\section*{Acknowledgments}
I thank O. Bergman for useful discussions.
This work  is supported in part by the
US-Israel Binational Science Foundation grant no.~2000359.

%%%%%%%%%%%%%%%%%%%%%%%%%%%%%%%%%%%%%%%%%%%%%%%%%%%%%%%%%%%%%%%%%%%%%%%%%%%%%%%%

\end{document}